\documentclass[]{elsarticle}
\usepackage{amsmath}
\usepackage{graphicx}
\usepackage{csquotes}
\usepackage{float}
\usepackage{caption}
\usepackage{subcaption}
\usepackage{hyperref}

\begin{document}
\title{Physical Properties of the Spin Hamiltonian on Honeycomb Lattice Samples with Kekul\'e and Vacuum
 Polarization Corrections}

\author[cbpf,iftm]{Ricardo Spagnuolo Martins\corref{cor1}}
\ead{ricardomartins@iftm.edu.br}
\cortext[cor1]{Corresponding author}
\author[ifsud]{Elena Konstantinova}
\ead{elena.konst@ifsudestemg.edu.br}
\author[ufes]{Humberto Belich}
\ead{humberto.belich@ufes.br}
\author[cbpf]{Jos\'e Abdalla Helay\"el-Neto}
\ead{helayel@cbpf.br}

\address[cbpf]{Centro Brasileiro de Pesquisas F\'isicas, 150 Xavier Sigaud
  Street, Rio de Janeiro, RJ, Brazil}
\address[iftm]{Instituto Federal de Educa\c c\~ao Ci\^encia, e Tecnologia do
  Tri\^angulo Mineiro, km 167 MG 188 Highway, Paracatu, MG, Brazil}
\address[ifsud]{Instituto Federal de Educa\c c\~ao Ci\^encia, e Tecnologia do
  Sudeste de Minas Gerais, 1283 Bernardo Mascarenhas Street, Juiz de Fora, MG,
  Brazil} 
\address[ufes]{Universidade Federal do Esp\'irito Santo, 514 Fernando Ferrari
  Av., Vit\'oria, ES, Brazil}

\date{\today}

\begin{abstract}
  Magnetic and thermodynamical properties of a system of spins in a
  honeycomb lattice, such as magnetization, magnetic susceptibility
  and specific heat, in a low-temperature regime are investigated by
  considering the effects of a Kekul\'e scalar exchange and QED vacuum
  polarization corrections to the interparticle potential. The spin
  lattice calculations are carried out by means of Monte Carlo
  simulations. We present a number of comparative plots of all the
  physical quantities we have considered and a detailed analysis is
  presented to illustrate the main features and the variation profiles
  of the properties with the applied external magnetic field and
  temperature.
\end{abstract}


\maketitle

\section{Introduction}
\label{sec:introduction}
Graphene materials are of a great deal of interest to scientists and
engineers from all fields by virtue of their unusual and important
properties\cite{Geim2009a,Geim2007a,Geim2007b,Katsnelson2007a}. The
atoms in graphene are arranged in one flat honeycomb (or hexagonal)
lattice, so we investigated the effects of some topological
deformations in a honeycomb lattice to shed some light on the
influence of the topology of the honeycomb lattice that could
possibily be extended to the discussion in graphene samples. In
particular, the properties of graphene and graphene-like materials
have been deeply inspected from both experimental and theoretical
approaches by means of different methods (e.g. see
\cite{Katsnelson2012a,Pati2011a,Katsnelson2007b,Novoselov2005b,Vozmediano2010a}
for recent reviews.)

One of the most interesting aspects of graphene, from a theoretical point of
view, is the close relation with Quantum Field Theory
(QFT)\cite{Semenoff1984a,Haldane1988a,Gonzalez1993a,Marino2015a}. Such a
connection has arisen due to the fact that the dispersion relation is linear
near the Dirac points (also called "valleys", and are the points where the
energy is zero). This behavior leads to the appearance of low-energy excitations
described by a 2-dimensional massless Dirac equation, with a much lower
counterpart of the speed of light, v (experimentally v $\sim$ c/300).  The
appearance of quantum anomalies in hexagonal network graphene-type systems is a
neat example of this connection \cite{Semenoff1984a}. There has been a number of
studies addressing fundamental questions of QFT-quantum anomalies
\cite{Semenoff1984a,Haldane1988a}, and more recently on the relationship between
QFT and fractional fermion number
\cite{Hou2007a,Jackiw2007a,Seradjeh2008a,Weeks2010a,Obispo2014a,Niemi1986a}.

Real graphene surfaces are not perfectly smooth and their physical properties
depend on the geometry the deformations, lattice imperfections and also whether
the structure has a hollow form, like a sphere, ellipsoid or a tube
(i.e. fullerenes and carbon nanotubes)
\cite{Vozmediano2008a,Gonzalez1993a,Georgiou2011a}.

In the work \cite{Chamon2000a}, Chamon discusses the possibility of describing
the curvature of a carbon nanotube with a U(1) continuous symmetry, and its
implications, related to the Kekul\'e distortions in the graphene plane. These are
natural oscillations of the carbon bond lengths simultaneously stretching and
compressing in alternating bonds. His results were supported by
\cite{Krotov1997a} at the time. In Ref. \cite{Hou2007a,Chamon2008a,Chamon2008a},
this idea is developed in further details. In the work \cite{Jackiw2007a},
Jackiw and Pi describe how a distortion in a lattice (called Peierls'
distortion) can be represented by a coupling of the Dirac field to a massive
scalar field, which is a measure of the lattice distortion. They associate this
scalar field to the Kekul\'e texture, as shown in the previous works of Chamon
et al. in order to develop a chiral gauge theory of graphene.

From the studies on vortex formation in \cite{Hou2007a}, it comes out that a
chiral gauge theory for graphene \cite{Jackiw2007a} presents a spinor structure
of fermionic zero-mode (fermionic excitations zero energy), which is not
modified by the addition of a gauge vector potential chiral coupled with
fermions. In fact such coupling promote the idea of magnetic fields fictitious
in graphene. This has yielded very fruitful consequences both experimentally and
theoretically in the description of elastic deformations and the study of the
formation of topological defects with its influences on the electronic
properties of graphene
\cite{Meyer2007a,Cortijo2007a,Juan2007a,Sitenko2007a,Guinea2008a,Kim2008a,Levy2010a,Juan2011a,Vozmediano2010a,Cortijo2007b,Fujita2011a}

In experimental scenarios, it is possible to observe (by Landau levels
measurements) intense pseudomagnetic field (up to 300T) due to tensions in
graphene \cite{Levy2010a}. It is possible to measure, through a microscope of
scanning tunneling, Aharonov-Bohm interference due to local deformations in the
network \cite{Juan2011a}. A detailed study of gauge fields in graphene can be
found in \cite{Vozmediano2010a}, in a approach of elastic deformations, the
emergence of topological defects in a curved environment is represented by a
low-energy Hamiltonian for graphene and some of its effects on electronic
properties are discussed \cite{Cortijo2007b,Fujita2011a}.

We have explored the idea that these surface imperfections can be described by a
scalar massive field and propose that this scalar is a massive propagating
degree of freedom that couples electrons and yields an effective interaction,
which is spin-dependent and exhibits a screening parameter, $\xi$, that is the
mass of the exchanged boson which we call from now on Kekul\'e particle.

In this work, we have simulated the physical properties (specific
heat, magnetization and magnetic susceptibility) at low temperatures
in a honeycomb spin lattice under the influence of the potentials
given by the electromagnetic interaction with loop correction and also
due to the exchange of the Kekul\'e boson, in order to account for the
non-smoothness of the honeycomb lattice surface. To realize this property, we
obtained non-relativistic interaction potentials given by the exchange
of a Kekul\'e particle in a quantum field-theoretic description. The
procedure to obtain such potentials from a Feynman diagram was
initially stated by Sucher and Feinberg
\cite{Sucher2008a,Sucher2007a,Feinberg1989a,Sucher1989a}. Essentially,
the method consists of examining a scattering via Feynman diagrams and
performing a low relativistic approximation on the scattering
amplitude. It is shown that this procedure yields the correct
interaction potentials previously known, including Coulomb interaction
from electron-electron QED scattering, first-order relativistic
corrections (known as Breit potential and Gaunt-Moller potential) and
Yukawa potential. This method is explored in further details in the
works by Dobrescu and Mocioiu \cite{Dobrescu2006a} and enhanced in the
following works \cite{Ferreira2015,Malta_2016}.

By following this path, we are able to obtain the interaction potentials
necessary to simulate the physical properties of graphene that we wish to
explore considering the electron-electron scattering mediated by the exchange of
a Kekul\'e boson. We also compared our results with the works of
\cite{Pei-Song2010,He2011}, where the specific heat and susceptibility of
graphene were analyzed theoretically using renormalization group and gaussian
correction methods for low temperatures.

The paper is organized as follows. In Section \ref{sec:methods}, we describe our
method of calculations and state the Hamiltonian used in our simulations. Also,
we describe in details all the computational methods, the variables and the
units that were used. In Section \ref{sec:results-discussions}, we present the
results of our calculations and discuss their general aspects. Finally, Section
\ref{sec:conclusions}, we discuss the details and features of our results,
present our conclusions and discuss positive and negative aspects of our
attempts and the limitations of the simulations we have carried out.

\section{Method of calculation}
\label{sec:methods}

Our model consists of a flat surface square honeycomb lattice
  of 288 sites. Each site contains a non-itinerant spin, or a vector,
  that is free to rotate through a fixed point continuously throughout
  all possible values in a solid angle of 4$\pi$ but not move among
  lattices. In other words, we allowed the point of the vector to
  rotate as if it was inside a sphere, assuming possibly every point
  in the surface of this sphere. The curvature of the surface is
  modeled in the interaction via the Kekul\'e scalar. In each site,
  the magnetic moment interact with its neighboring sites by
  Heisenberg-type and dipole-dipole-type interactions given by the
  Hamiltonian \eqref{eq:1} and also by a vacuum polarization
  correction potential.

In regard to the issue of quantum interactions among classical
  agents, we would like to state that we try in this paper to propose
  and analyze new kind of interaction potentials. These potentials
  were obtained by an underlying more fundamental physical
  interactions between fundamental particles. Despite that, these are
  long range potentials that can reach macroscopic extents and could
  be present among interaction of classical spins. This is in the
  heart of the method. The same process of obtaining these kind of
  potentials described in previous references can also result in
  classical Coulomb potentials, for example, when considered the
  quantum interactions of electrons mediated by a photon propagator.

A similar approach was used another recent reference \cite{Zheng2012a} in which
  they perform an experimental investigation of the contribution of
  spin interactions of nucleons inside a nucleus of $^3 He$, which was
  theoretically proposed in the same fashion and actually share some
  of the same references with our work, concerning the process of
  obtaining these long range interaction potentials.

We have used the Monte Carlo method with the Metropolis algorithm
\cite{Metropolis1953a,Landau2014a,Binder1996a}. It was chosen due to its
capability to obtain an equilibrium macro-state of a physical system with many
coupled degrees of freedom (such as cellular Potts Models and its
generalizations, like the Heisenberg model) at a given temperature T. We chose
an initial micro-state and performed a very large number of random
transformations by a deterministic procedure until we achieve an equilibrium
macro-state. The initial micro-state was a parallel array of spins arranged in
the sites of a honeycomb lattice. We then randomly changed this configuration and
evaluated the change of the overall energy of this mew micro-state and compared
to the previous configuration. If $\Delta E > 0$, this micro-state replaces the
previous one with a probability $e^{-\Delta E/k_bT}$. In a preliminary
simulation, we evaluated that the number of steps required to reach the
equilibrium state was in the order of $n=10^4$ for each site in the lattice and
we used this number of steps in all of our calculations. The final state
corresponded to the stable configuration and was interpreted as the equilibrium
macro-state.

Using the method described above, we obtained spin configurations, thermal
equilibrium magnetization, magnetic susceptibility and specific heat of the
chosen structure. The Hamiltonian for the computed system used was

\begin{align}
  \label{eq:1}
  H= &-\vec{B} \cdot \sum\limits_{<i,j>} \vec{S}_i - \left( J\sum\limits_{<i,j>}
       \vec{S}_i\cdot\vec{S}_j +  \tilde{J} \sum\limits_{<i,j>}\frac{e^{-\xi r_{ij}}}{4\pi
       r_{ij}} \vec{S}_i\cdot \vec{S_j}\right)\nonumber \\  
  &- \omega \sum\limits_{i<j} \frac{(\vec{S}_{i}\cdot\hat{e}_{ij})(\vec{S}_j\cdot
    \hat{e}_{ij})-(\vec{S}_i\cdot\vec{S}_j)}{r_{ij}^{3}}\nonumber \\
     &- \tilde{\omega}   \sum\limits_{<i,j>} \frac{(3+3\xi
       r_{ij}+\xi^2r_{ij}^2)(\vec{S}_i\cdot\hat{e}_{ij})(\vec{S}_j\cdot\hat{e}_{ij})-(1+\xi
       r_{ij})(\vec{S}_i\cdot\vec{S}_j)}{4\pi m^2 r_{ij}^3}\nonumber \\
     &+\frac{\alpha}{r_{ij}^2}\sum\limits_{i<j} \left( 1+\frac{\alpha}{4\pi}\frac{e^{-2mr_{ij}}}{(mr_{ij})^{3/2}} \right).
\end{align}

The first term in the first line, the summation represents the coupling of spins
to an external magnetic field B. The terms inside the parenthesis represents the
ferromagnetic exchange between the nearest neighbors due to the exchange of a
photon and the Kekul\'e scalar respectively, with coupling constants $J$ and
$\tilde{J}$. The second line is the dipole-dipole interaction
due to the electromagnetic interaction with strength $\omega$. The third
line stands for the dipole-dipole interaction due to the
exchange of the scalar boson with strength $\tilde{\omega}$. The fourth line
represent the second order electromagnetic loop correction. The $\vec{S}_i$'s
are the three-dimensional magnetic moments of unit length; $\hat{e}_{ij}$ stands
for the unit vectors pointing from the lattice site $i$ to the lattice $j$ and
$r_{ij}$ represent the distances between these lattice sites. The quantities
$\omega$ and $\tilde{\omega}$ may be regarded as the coupling constants for the
exchange term and the dipole-dipole interaction respectively. The parameter
$\xi$ is the mass of the scalar boson and $m$ is the mass of the electron.

We assume that the external magnetic field is orthogonal to the plane of the
structure. The energy and the applied magnetic field are expressed in units of
the coupling constant $J$. The temperature is expressed in the units of $J/k_b$,
where $J$ is the magnitude of the coupling constant and $k_b$ is the Boltzmann
constant.

We obtain the magnetic susceptibility $\chi$ (in this case along the OZ-axis), by
using the Monte Carlo method, according to the expression
\begin{equation}
  \label{eq:2}
  \chi=\frac{1}{k_bTN}\left( \langle m_z^2\rangle - \langle m_z\rangle^2 \right),
\end{equation}
where N is the number of spins in the system and $\langle m_z \rangle$ is the
mean magnetization per spin in the z-direction. the specific heat C is obtained
from the energy fluctuations relation
\begin{equation}
  \label{eq:3}
  C=\frac{1}{k_bT^2N} \left( \langle E^2\rangle - \langle E\rangle^2 \right)
\end{equation}
where $\langle E\rangle$ is the mean energy per spin. For calculating the
specific heat we used $B=0$.

\section{Results}
\label{sec:results-discussions}

After the general presentation of the Hamiltonian model we have adopted to
pursue our investigations, we can from now start off the presentation of the
calculations and the corresponding plots for the physical properties we are
interested in, namely, the magnetization (Subsection \ref{sec:magnetization}),
the magnetic susceptibility (Subsection \ref{sec:susceptibility}) and the
specific heat (Subsection \ref{sec:specific-heat}).

Since there shall be shown many graphs, we have made the option to cast our
comments and general discussions of the results we have found in the final
Section (Section \ref{sec:conclusions}) of our paper.

\subsection{Magnetization}
\label{sec:magnetization}

We show the results for the calculation of the magnetization in terms of the
applied magnetic field for different values of $\xi$ 

\begin{figure}[H]
  \centering
  \includegraphics[]{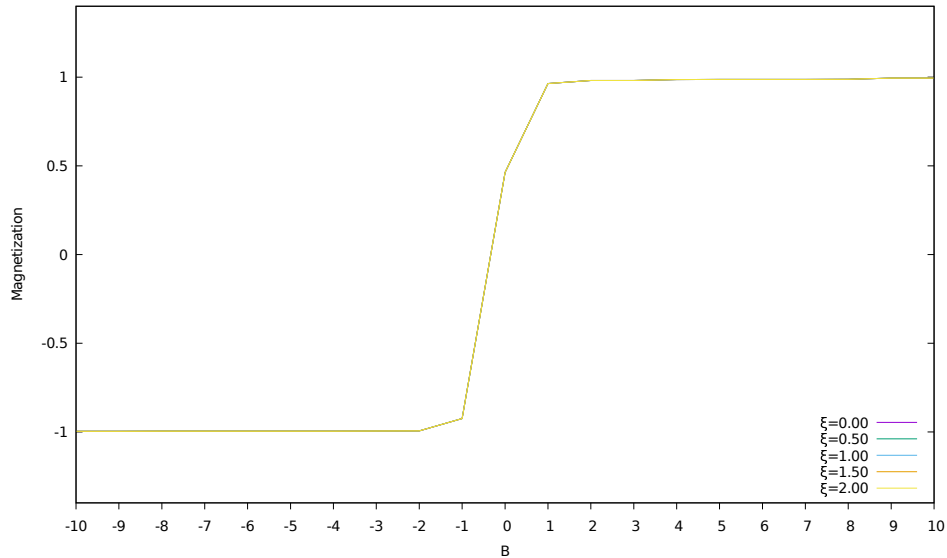}
  \caption{Plots of magnetization versus applied field for all values of $\xi$
    for Heisenberg term only. (color online)}
  \label{fig:mag1}
\end{figure}

In this graph, we calculated the magnetization considering only the Heisenberg
interaction. Notice that since there is no exchange of the Kekul\'e particle, we
see only one magnetization curve. There is a full magnetization at about an
external field strength of $\pm 2$, meaning that all spins are aligned. There is a
positive magnetization of $0.46$ in the absence of the external field and no
magnetization at a field value of around $-0.3$.

\begin{figure}[H]
  \centering
  \includegraphics[]{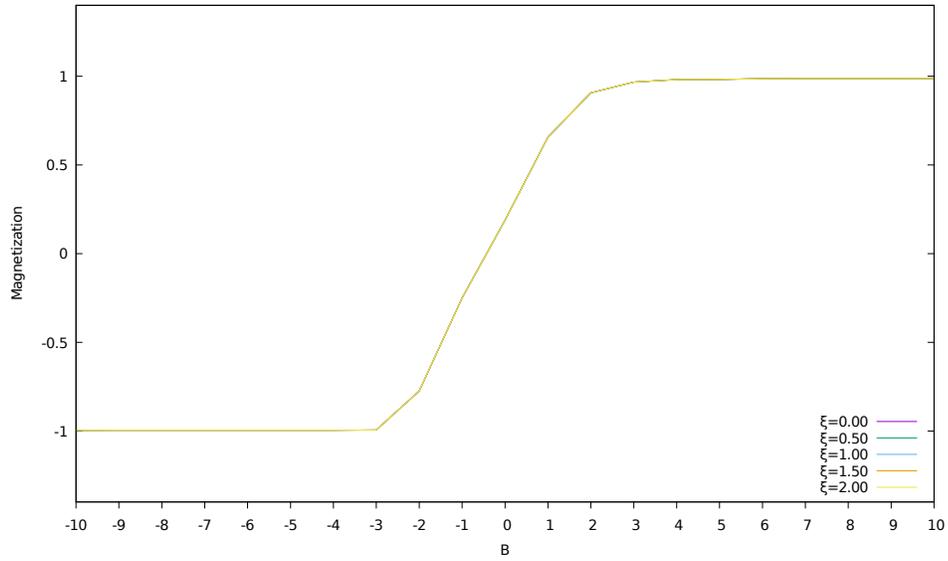}
  \caption{Plots of magnetization versus applied field for all values of $\xi$,
    with dipole term included. (color online)}
  \label{fig:mag2}
\end{figure}

When we include the dipole interaction, we see that the full magnetization
occurs at field values of $-3$ and $4$. In the absence of the external field,
the magnetization is $0.19$ and there is no magnetization at a field value of
$-0.5$.

\begin{figure}[H]
  \centering
  \includegraphics[]{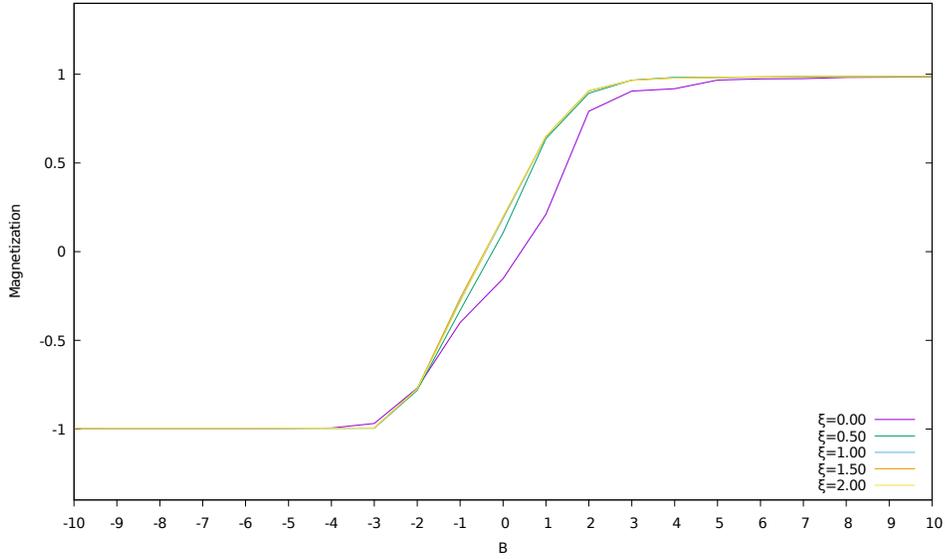}
  \caption{Plots of magnetization versus applied field for all values of $\xi$,
    with first Kekul\'e term included. (color online)}
  \label{fig:mag3}
\end{figure}

As we include the first Kekul\'e interaction term, we clearly see a distinct
behavior for the magnetization curve of different masses of the Kekul\'e
boson. For higher values of $\xi$, the magnetization pattern is exactly like
\ref{fig:mag2}. We bring the attention to the magnetization curve for
$\xi=0$. It has a magnetization of $-0.25$ in the absence of the external field
and zero magnetization for field values of $0.4$. For the next higher value
calculated, $\xi=0.5$, we see a behavior only slightly different from higher
values of $\xi$.

\begin{figure}[H]
  \centering
  \includegraphics[]{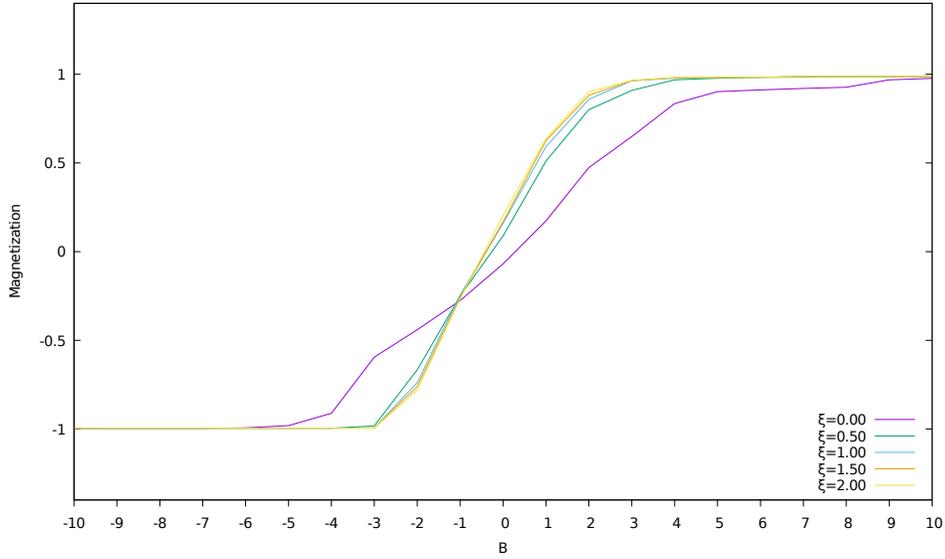}
  \caption{Plots of magnetization versus applied field for all values of $\xi$,
    with second Kekul\'e term included. (color online)}
  \label{fig:mag4}
\end{figure}

In this calculation, we have all the previous interaction terms and have
included the second Kekul\'e term. The $\xi=0$ line shows that for a massless
Kekul\'e particle, the material becomes much less responsive to external magnetic
fields. We see a clearer differentiation between the higher $\xi$ curves, but
they maintain the same behavior as in previous calculations. In order to become
fully magnetized for $\xi=0$, we need to apply a field twice as high as for
higher $\xi$ in both orientations.

\begin{figure}[H]
  \centering
  \includegraphics[]{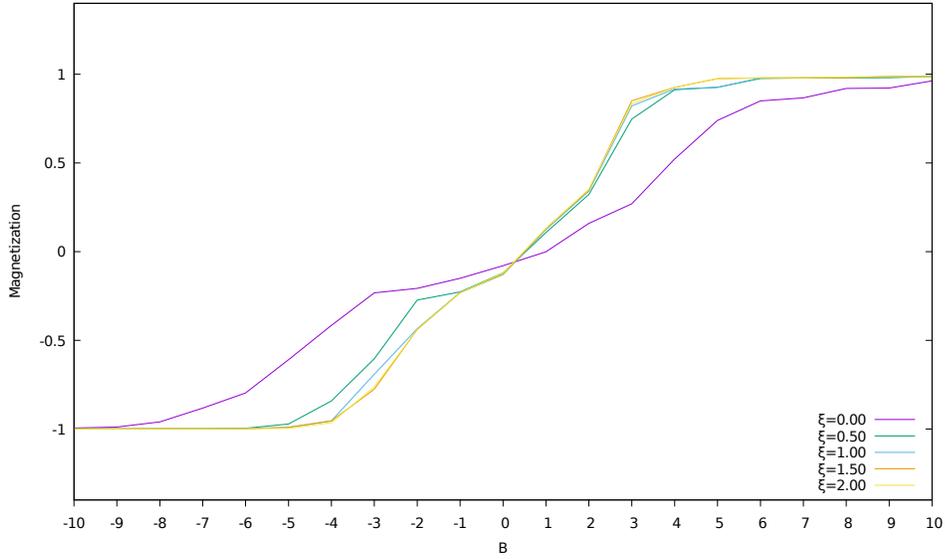}
  \caption{Plots of magnetization versus applied field for all values of $\xi$,
    with loop correction included. (color online)}
  \label{fig:mag5}
\end{figure}

In this calculation, we have included the loop correction term to the previous
calculations. The influence of this interaction is clear mainly in the
responsiveness to the external magnetic field. All curves need about twice the
strength of the external field in order to achieve a full magnetization. Also,
all the curves show a negative magnetization in the absence of external fields
of $-0.12$ for $\xi\neq 0$ and $-0.08$ for $\xi=0$.

\subsection{Susceptibility}
\label{sec:susceptibility}

Here we present the results of the calculations for the magnetic susceptibility
for all values of $\xi$.

\begin{figure}[H]
  \centering
  \includegraphics[]{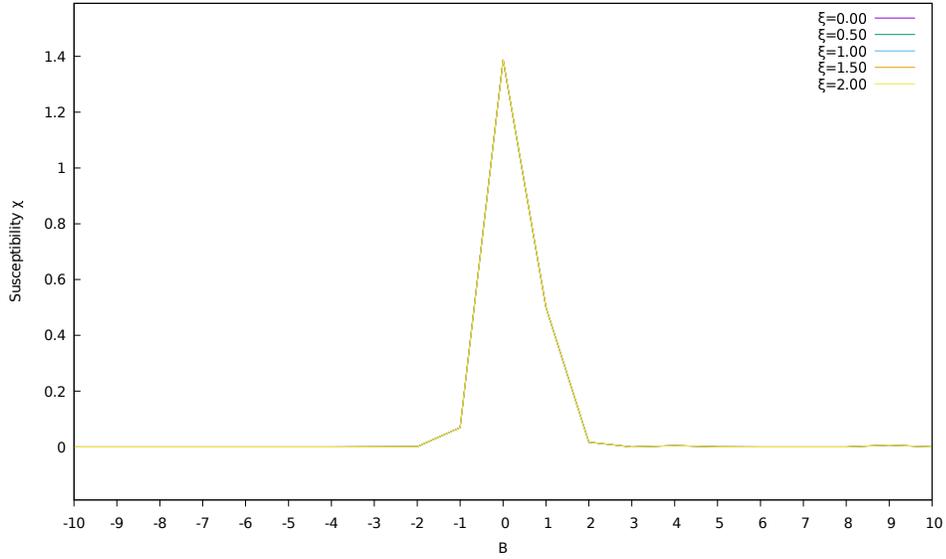}
  \caption{Plots of the magnetic susceptibility versus applied field for all
    values of $\xi$, with the Heisenberg interaction. (color online)}
  \label{fig:sus1}
\end{figure}

We see the highest magnetic susceptibility at $B=0$, with a small asymmetry for
negative and positive external fields, but vanishing quickly for field values of
$|B|> 2$.

\begin{figure}[H]
  \centering
  \includegraphics[]{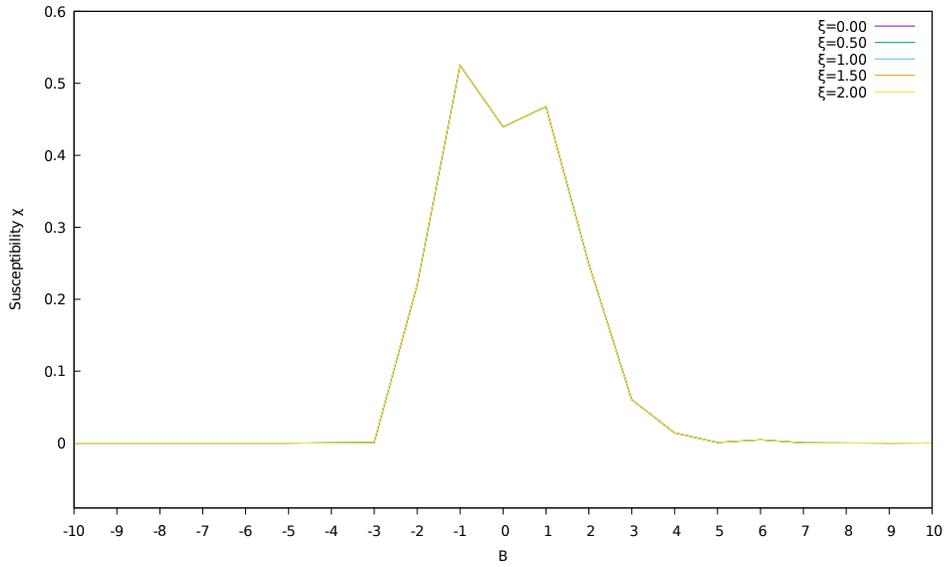}
  \caption{Plots of the magnetic susceptibility versus applied field for all
    values of $\xi$, with the dipole interaction included. (color online)}
  \label{fig:sus2}
\end{figure}

When included the dipole interaction term, the results show now 2 peaks of
magnetic susceptibility, being higher for negative fields. Also, notice that the
strength of the susceptibility decreases 3 times in comparison to
\ref{fig:sus1}.

\begin{figure}[H]
  \centering
  \includegraphics[]{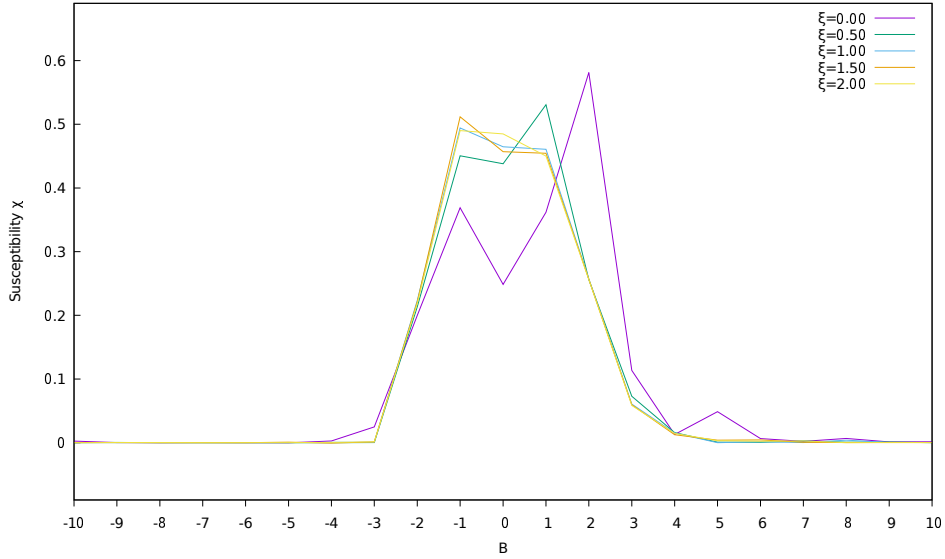}
  \caption{Plots of the magnetic susceptibility versus applied field for all
    values of $\xi$, with the first Kekul\'e term included. (color online)}
  \label{fig:sus3}
\end{figure}

Here we have included the first Kekul\'e interaction term to the previous
calculations. We clearly see a pattern of asymmetry for the peaks of
susceptibility for positive and negative fields, being different for all values
of $\xi$. For the lower values of mass $\xi=0$ and $\xi=0.5$, we see that the
system responds more strongly to positive fields than negative fields, but that
response comes with the cost of applying a stronger positive field than a
negative field, e.g., the peak occurs at $B=2.1$ and $B=-1.0$ for $\xi=0$. Also,
we see a third small peak at $B=5$ for a massless boson.

For the higher values of mass $\xi=1.0$ and $\xi=1.5$, we see the pattern is
reversed in relation to the previous lower masses. The lowest peak of magnetic
susceptibility is seen for the highest values of $\xi$, while at $\xi=2.0$ we
can see only one diffuse peak at $B=-1$ that drops sharply for $B>1.0$.

\begin{figure}[H]
  \centering
  \includegraphics[]{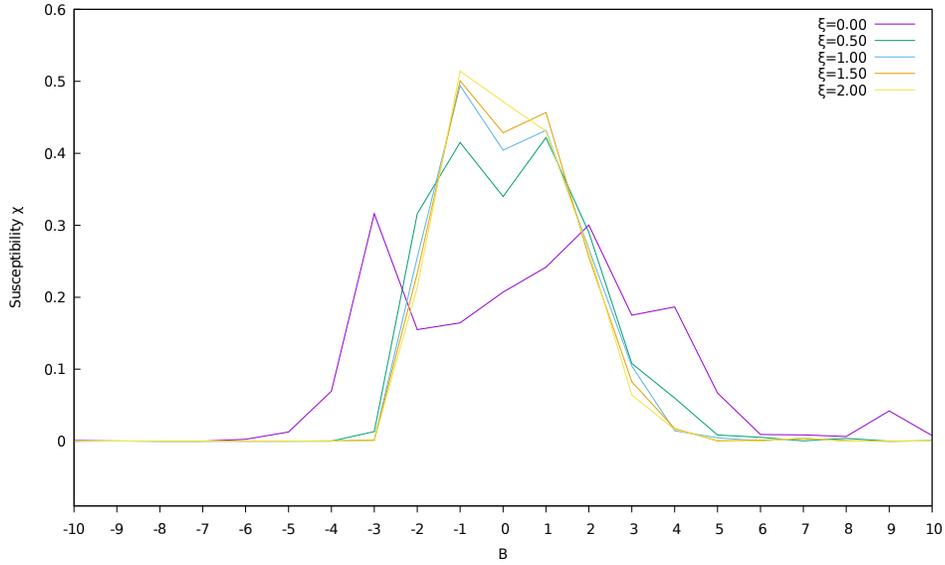}
  \caption{Plots of the magnetic susceptibility versus applied field for all
    values of $\xi$, with the second Kekul\'e term included. (color online)}
  \label{fig:sus4}
\end{figure}

Here we have include the second Kekul\'e interaction term to the previous
calculations. Now the lowest susceptibilities are seen with the lower values of
$\xi$, a behavior opposite from \ref{fig:sus3} with a prevalence for negative
fields. Again, we see for the highest value of $\xi=2.0$ what seems to be only
one peak, in contrast to 2 peaks for all other values of $\xi$ and what seems to
be even 3 peaks for $\xi=0$.

\begin{figure}[H]
  \centering
  \includegraphics[]{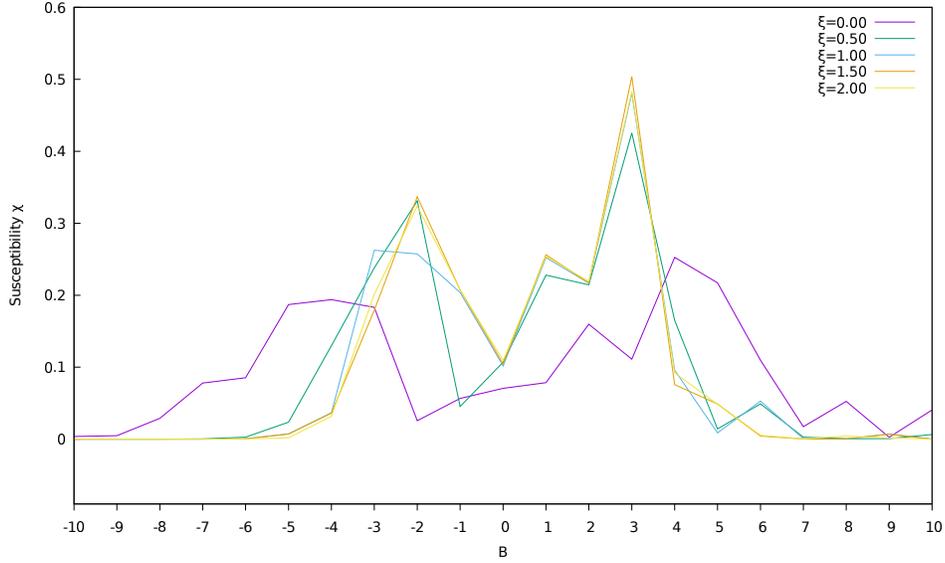}
  \caption{Plots of the magnetic susceptibility versus applied field for all
    values of $\xi$, with the loop correction term included. (color online)}
  \label{fig:sus5}
\end{figure}

In this calculation we have included the loop correction to the previous
calculations. Again, we see a shift in the pattern of peaks, now favoring the
positive field values, opposite of \ref{fig:sus4}. We bring the attention that
with the inclusion of the loop correction, the curve for $\xi=2.0$ now became
very similar to the $\xi=1.5$ curve, showing 2 distinct peaks.

Also, we point out that in \ref{fig:sus4}, we see for $\xi\neq 0$ that the peaks
occur to very low values of the external fields, about $\pm 1$. In
\ref{fig:sus5}, both peaks occur at much higher values of the external field,
$B=-2$ and $B=3$.

\subsection{Specific Heat}
\label{sec:specific-heat}

In this section, we present the results of the calculations for the specific
heat for all values of $\xi$.

\begin{figure}[H]
  \centering
  \includegraphics[]{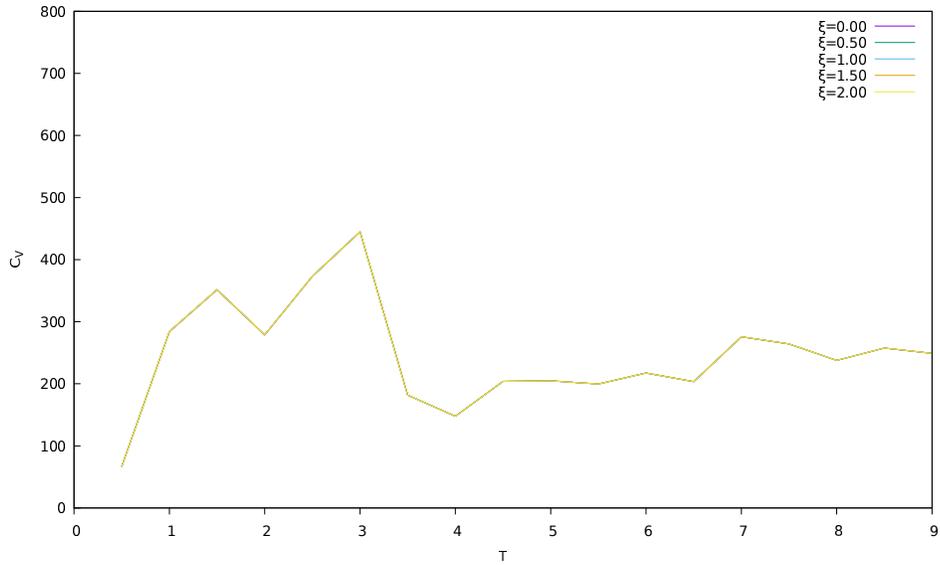}
  \caption{Plots of the specific heat versus temperature for all values of
    $\xi$, with only the Heisenberg interaction. (color online)}
  \label{fig:cv1}
\end{figure}

We see here the results for the specific heat when calculated with only the
Heisenberg interaction term. We see a domain of a high specific heat, possibly
with two peaks, in the temperature region from 0 to 4. From this temperature on,
we see a tendency of increasing specific heat.

\begin{figure}[H]
  \centering
  \includegraphics[]{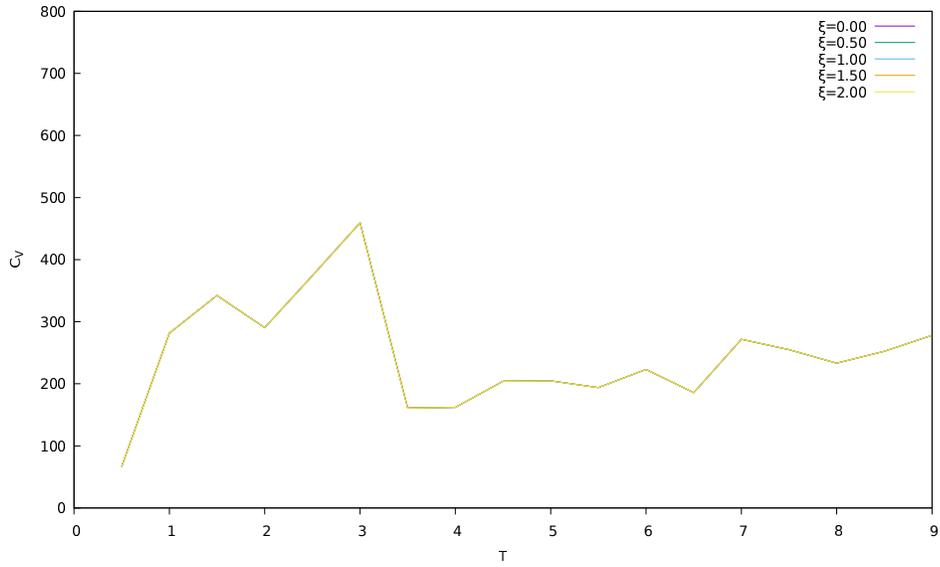}
  \caption{Plots of the specific heat versus temperature for all values of
    $\xi$, with the dipole interaction included. (color online)}
  \label{fig:cv2}
\end{figure}

In figure \ref{fig:cv2}, we have included the dipole interaction term. We see
only small changes from \ref{fig:cv1}, barely visible, that can be attributed to
the randomization procedure of the Monte Carlo method.

\begin{figure}[H]
  \centering
  \includegraphics[]{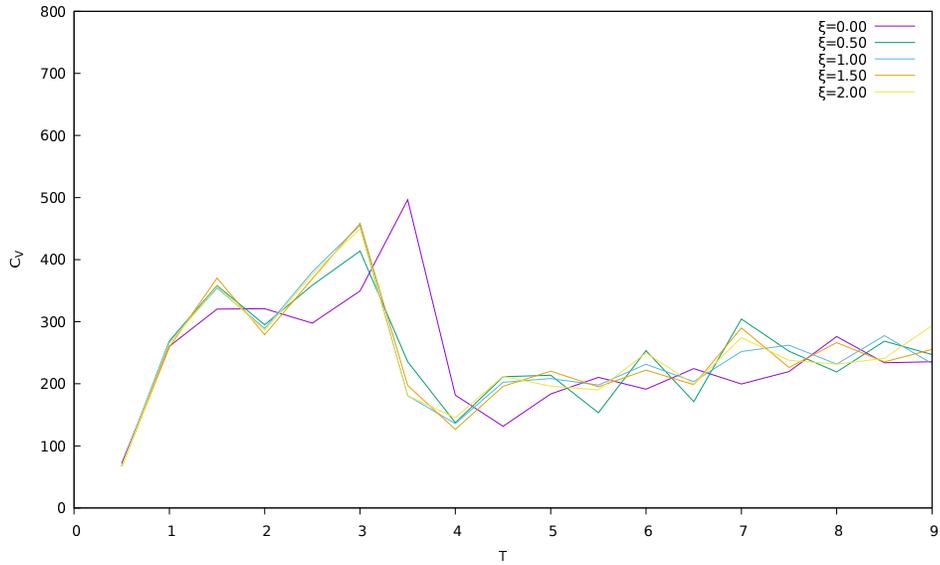}
  \caption{Plots of the specific heat versus temperature for all values of
    $\xi$, with the first Kekul\'e term included. (color online)}
  \label{fig:cv3}
\end{figure}

Here we have included the first Kekul\'e term to the previous calculations. Now we
can see a difference in behaviour from \ref{fig:cv1} and \ref{fig:cv2}. The
curve for $\xi=0$ is shifted to the right in relation to the other $\xi\neq 0$
lines. Also, for $\xi=0$ we have a clear distinctive second peak in the specific
heat in relation to the first peak, while for $\xi\neq 0$ we have what looks
like two peaks in the region $T=[0;4]$. For all lines, we have the same
increasing specific heat for $T>4$, with stronger oscillations.

\begin{figure}[H]
  \centering
  \includegraphics[]{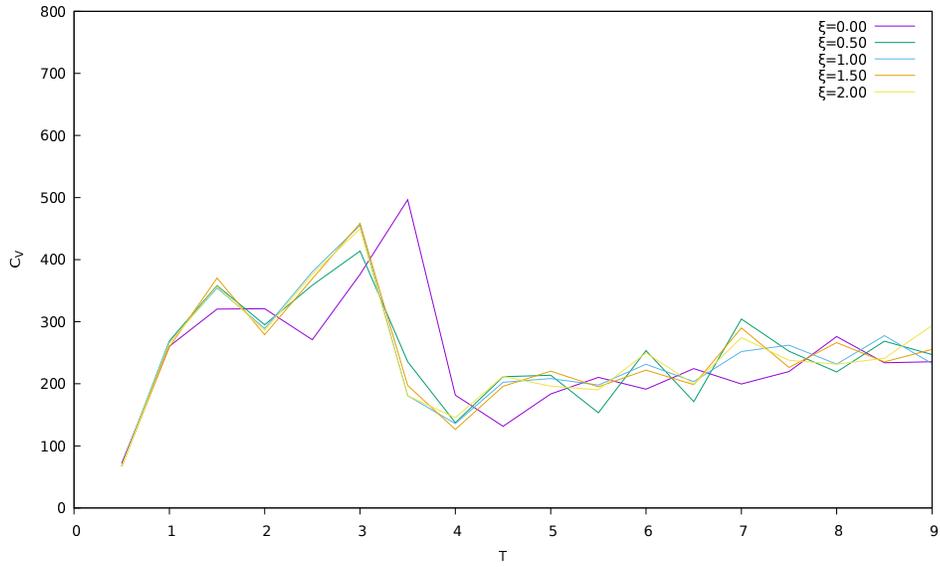}
  \caption{Plots of the specific heat versus temperature for all values of
    $\xi$, with the second Kekul\'e term included. (color online)}
  \label{fig:cv4}
\end{figure}

Here we have included the second Kekul\'e term to the previous calculations. We
see only small differences from the previous calculation \ref{fig:cv3} and we
assume that they are consistent with fluctuations from the Monte Carlo method.

\begin{figure}[H]
  \centering
  \includegraphics[]{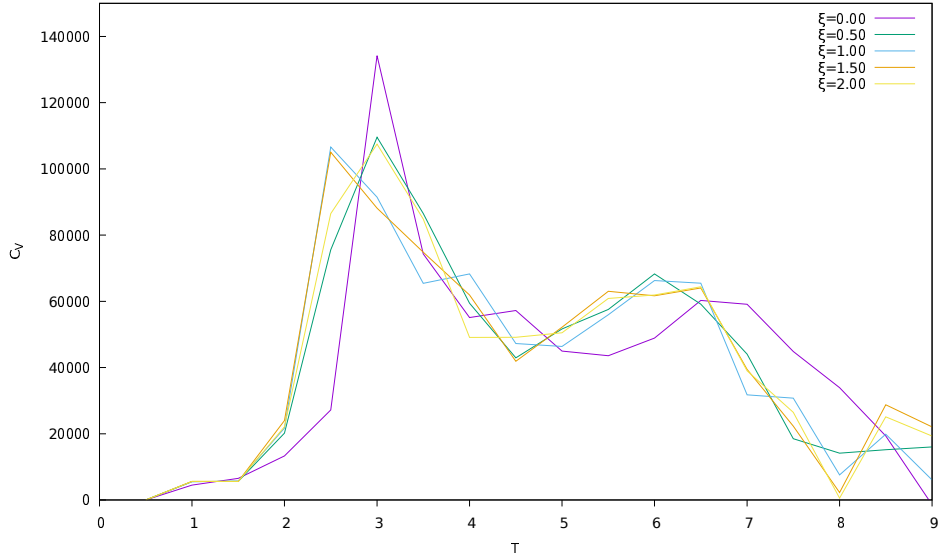}
  \caption{Plots of the specific heat versus temperature for all values of
    $\xi$, with the loop correction term included. (color online)}
  \label{fig:cv5}
\end{figure}

In figure \ref{fig:cv5}, we have include the loop correction term to the
previous calculations. This have a great influence on the behavior of the
specific heat. We see now only one small, distinctive peak in the region
$T=[0;4]$, arising exponentially from the lower values of $T$. In the region
$T>4$, we see only an elevation in the line, which then has a decreasing
pattern. This contrasts with the increasing pattern in the same region in the
previous calculations.

We would like to call the attention to the change in the vertical scale in
figure \ref{fig:cv5}. It states that the values of the specific heat are 233
times larger than in previous calculations, indicating a strong influence of the
loop correction term to the specific heat.

This is in agreement with \cite{Pei-Song2010}, which predicts a specific heat
for low temperatures proportional to $C_v \propto T^2/|\ln{T^2}|$ with second
order loop corrections.

\section{Discussion and final considerations}
\label{sec:conclusions}

We have shown that our proposed method of obtaining interaction
potentials from quantum field theory and using them to simulate
physical systems provide results that can be tested against experiment
or compared to other theoretical methods of evaluation. Also, the idea
that we explored that the surface curvature of the honeycomb lattice
can be modelled by a scalar particle, which we called Kekul\'e boson,
has show to provide differences in the calculated properties that can
be looked for in experiments.

In Section \ref{sec:magnetization}, we have shown that the mass of the Kekul\'e
particle influences the interaction with the external magnetic field. We see in
all the calculations that the higher the mass of the Kekul\'e boson, the stronger
the interaction with the external field is, given all the possible
configurations. Since the high mass of the Kekul\'e boson is related to the
curvature of the lattice plane, we conclude that higher curvatures or defects
in the planarity of the lattice make the material more subject to the
influence of external magnetic fields.

We also see that there is apparently a limit to the value of $\xi$, or in other
words the curvature, can be. We see in figures \ref{fig:mag3}, \ref{fig:mag4}
and \ref{fig:mag5} that for values of $\xi\geq 1.5$, the difference in the
magnetization curve is practically non-existent among those lines, indicating
that if we increase the curvature further (i.e. increase the mass), we would
have no additional changes to the interaction with the external field.

In relation to the contributions of the individual terms calculated, we conclude
that the greatest contribution to the magnetization of the system comes from the
the dipole-dipole interaction, aside from the mass $\xi$. When this term is
introduced in the calculations (Fig. \ref{fig:mag2} and \ref{fig:mag4} the spins
in each lattice interact with each other in such a way that the external field
required to promote a full alignment of the spins in the system is 50\%
greater. As the external field interacts with each site of the lattice, in
return each lattice site interact with each other in such a way that the more
energetically stable state is the one in which all spins are perpendicular to
each other and not parallel. Thus, when this interaction is turned on in our
computations, we see an immediate increase in the required external field
necessary to align all the spins parallel to each other and also to the external
field.

We have shown that in our model, the calculations without the Kekul\'e
particle resulted in a slightly self-magnetic material, with a
positive ordering of spins without the influence of an external
magnetic field (Fig. \ref{fig:mag1} and \ref{fig:mag2}). When the
Kekul\'e terms were included in the calculation (Fig. \ref{fig:mag3}
and \ref{fig:mag4}), there was also an ordering of the spin states,
but the direction depended on the mass value of the Kekul\'e
particle. For low values of $\xi$, it showed an overall orientation in
the negative direction and for higher values it showed an overall
orientation in the positive direction. We could not probe for which
value of $\xi$ when the switch from negative to positive occurs,
neither for which value there is zero magnetization due to a
limitation in our implementation of the algorithm. Our results when
included the loop correction term (Fig. \ref{fig:mag5}) showed only
negative orientation of spins for all values of $\xi$, meaning that
these second order correction terms may have a strong influence in the
ordering of spins in the lattice. Since graphene materials have also
honeycomb lattices, we include some references in relation to the
highly debated question of intrinsic magnetic properties of graphene
(e.g. see the references
\cite{Katsnelson2012a,Makarova2006a,Esquinazi2005a,Esquinazi2002a,Hoehne2002a,Makarova2001},
to which our calculations propose alternative scenarios.

For this type of calculation, our algorithm swept from higher values
of the external magnetic field to the lower values in constant
steps. We also think that this method for probing the magnetization of
the sample could possibly show a characteristic of a hysteresis curve
for the magnetization, and this could also explain the asymmetry for positive and negative values of the external field B in for the magnetization.

Our method consider a variety of possible scenarios and possible answers in each
case, but one other limitation from our code implementation is that we rely on
experimental input for the parameters used, therefore making it unable to
present definitive answers to such questions with high degree of certainty in
topics where there is a current debate in literature and experiment. On a
positive note, we see that our model is flexible enough to adapt itself in such
way that may contemplate the possible scenarios in the debate and give some
physical insight in the issue, considering that new interaction terms can be
quickly turned on or off in order to account for the physical description
inquired.

In the calculations of the magnetic susceptibility, we have shown that each term
added to the calculation (magnetic dipole-dipole, Kekul\'e and loop) changes the
susceptibility to an external field, corroborating with the previous results of
the magnetization. We see an overall asymmetry in the susceptibility due to
positive fields v. negative fields. This is seen both in the range of the
existing susceptibility as well as it's intensity. This seems to indicate a
preferential direction of application of the magnetic field. This seemingly
preferential direction depends directly on the mass of the Kekul\'e particle, when
it is included in the calculations, and also on the magnetic dipole-dipole
interactions.

As we have shown in Fig. \ref{fig:sus1}, the strongest susceptibility occurs for
very weak fields, both in the positive and negative directions. As we increase
the intensity of the applied magnetic field, the susceptibility quickly
decreases for negative values of the applied field and about 50\% slower for
positive values of the applied field before it goes to zero. When the
dipole-dipole interactions are turned on in Fig. \ref{fig:sus2}, we still see a
slower decrease in the magnetic susceptibility towards positive applied
fields. But there is a remarkable difference from only Heisenberg interaction:
it increases for weak fields before it decreases for stronger fields. It seems
that for weak applied fields, there is a tendency of the spins to respond
strongly to the field, as they tend get parallel to each other. This quickly
reaches a limit value as the sample increases in the magnetization (more spins
get parallel to each other), and to increase further the number of such parallel
spins, it must be applied a much stronger magnetic field.

This pattern remains when Kekul\'e terms are added in \ref{fig:sus3}. When the
Kekul\'e-dipole interaction term was added in \ref{fig:sus4}, this pattern becomes
extremely dependent on the value of $\xi$. For $\xi=0$, for weak fields we see
again a preferential direction towards positive applied fields. The
susceptibility increases as the positive field increases, but decreases for
negative fields. The spins align themselves faster in the positive direction
than in the negative direction. For $|B|=2$, we see a point of change. For
negative values of the field, the spins become aligned faster and for positive
values of the field, the rate of alignment of spins reaches a peak value. As we
increase the value of $\xi$, this quickly changes, becoming each time more
symmetric around weak fields. For $\xi=2$, we see another asymmetry
occurring. For weak fields, the rate of alignment of the spins increases for
negative applied fields, and decreases for positive applied fields, never
increasing in the latter and reaching a peak value in the former
case. Nevertheless, it still decreases slower for positive applied fields than
for negative applied fields. We conclude that the curvature of the lattice plane
sheet, or the degree of irregularities, have a strong influence in the magnetic
susceptibility of the sample.

When we included also the second order loop correction term in
Fig. \ref{fig:sus5}, we still see an asymmetry highly dependent on the value of
$\xi$ similar to \ref{fig:sus4}, but as we increase the value of $\xi$, the
magnetic susceptibility pattern tends to maintain the symmetry around weak
fields, being virtually the same for $\xi=1.5$ and $\xi=2.0$. We still see a
preference for positive applied fields for high $|B|$ values, but this is not
seen in weak $|B|$ values. This shows that the loop correction effects can be
significant when considering this property. In a disclaimer note, we admit to a
limitation in our analysis that we can't be sure if we have overestimated or
underestimated the loop correction term. The value of the parameter used was
obtained via energy optimization considerations and is by no means unique since
our calculations were not \textit{ab initio}. The parameters used for the other
terms were chosen from a very limited range established using the same
considerations used for the loop term, all showing the same characteristics,
therefore not posing the same concern for the loop term.

For the calculation of the specific heat, we have shown that the magnetic
dipole-dipole interaction does not influence this property, as can be seen in
Fig. \ref{fig:cv2} and Fig. \ref{fig:cv4}. This means that the interaction among
spins in the lattice, both via exchange of photons or the Kekul\'e boson, do not
influence in how the lattice stores thermal energy. In other words, the magnetic
interactions do not influence the lattice vibrations due to thermal
fluctuations.

We saw in Fig. \ref{fig:cv1} and \ref{fig:cv3} that the specific heat is
slightly dependent on the value of $\xi$. We see a clear distinction of two
regions in the calculations: one with 2 high peaks until a value of
approximately $T=4$ and another where the specific heat steadily increases with
some fluctuations for $T>4$. As we saw in Fig. \ref{fig:cv3}, the value of $\xi$
shifts these patterns sideways. As we increase the values of $\xi$, the patterns
shifts to the left, but we see that this quickly reaches a limit.  For
$\xi\geq 0.5$, there are only very small changes in the intensities of the peaks
but no shifting in their value for $T$. We also saw that for $\xi>0$, there is
also a higher definition of the first peak in comparison to $\xi=0$. We did not
probe for a higher resolution in the values of $\xi$ to determine the precise
value of $\xi$ where this occurs in this work. Although, we see that the
influence of the $\xi$ reaches a limit fast in the capacity to store thermal
energy, being greater for lower values of $\xi$ and low values of $T$.

What came as a surprise to us in our calculations was the importance of loop
corrections in determining the specific heat of the honeycomb lattice, as seen in
Fig. \ref{fig:cv5}. We see an exponential increase in the C$_\text{v}$ for low
values of $T$, which was not present in the absence of the loop term. This is
consistent with the works of Pei-Song et al. \cite{Pei-Song2010}. They
calculated the specific heat of graphene due to loop corrections for low
temperatures using the renormalization group. In their work, they predicted that
the specific heat should have a temperature dependence proportional to
$C_\text{v} \propto T^2/|\ln{T^2}|$, which we have reproduced in our model
calculations. We have also shown that there is a second and less sharp peak in
the specific heat for values of $T$ about two times larger than the first
peak. Both peaks have significant higher values than those shown in
Fig. \ref{fig:cv1} and \ref{fig:cv3}. We also saw that the steadily increasing
specific heat is absent in Fig. \ref{fig:cv5}, but this could be due to the
difference in scale of the previous calculations. We intend to broaden the range
of temperature in future works. We see that the value of $\xi$ has the same
peak-shifting property as seen in Fig. \ref{fig:cv3}.

The aforementioned work of Pei-Song et al. \cite{Pei-Song2010} also predicted
the dependence of the magnetic susceptibility for low values of temperature to
be $C_\text{v} \propto T/|\ln{T^2}|$. Our algorithm did not allow for such
calculations, and we could not test the authors' prediction, although we have
calculated this property and its relation to the applied external magnetic
field. We should be able to improve on the algorithm in order to allow for this
calculation as well.

The property of the specific heat has shown to have a very low interference from
the magnetic dipole-dipole interaction. Both with and without the Kekul\'e terms,
the inclusion of the dipole-dipole type interaction has shown to not change any
characteristics of this property. The Kekul\'e interaction alone was seen to have
some influence in the specific heat, mostly considering the calculated values
for the mass of the Kekul\'e scalar. Other than that, it does not alter the main
characteristics of this property in relation to the Heinsenberg interaction
alone. The parameter that mostly affects the specific heat is the loop
correction. It increases the value of the property by 233 times, single out one
peak value in contrast to two peak values in the $T=[0;4]$ region and shows a
second peak value in the $T>4$ region with a decreasing value of the property
for $T>7$. The very high peak value obtained in the $T=[0;4]$ region
corroborates with the results obtained by \cite{Pei-Song2010} using other
theoretical methods of evaluation for the same system.

The authors express their gratitude to the Brazilian Government's Funding
Agency, CNPq, for the invaluable financial support.


\end{document}